\documentclass{optica-article}

\journal{opticajournal} 

\articletype{Research Article}

\usepackage{lineno}

\begin{document}

\title{Heterodyne-Enhanced Sensing of Cavity Mode Spacing}

\author{Aaron D. Spector\authormark{1,*} and Laura Roberts\authormark{2}}

\address{\authormark{1}Deutsches Elektronen-Synchrotron DESY, 22603 Hamburg, Germany\\
\authormark{2}Max-Planck-Institut f\"ur Gravitationsphysik (Albert-Einstein-Institut) and Leibniz Universit\"at Hannover, 30167 Hannover, Germany}

\email{\authormark{*}aaron.spector@desy.de} 


\begin{abstract*} 
Characterizing the frequencies of resonant modes is critical to a wide variety of experimental applications. In this work we demonstrate a heterodyne-enhanced sensing scheme capable of precisely measuring the frequency spacing between longitudinal modes of a 19\,m optical cavity. The technique utilizes two resonant lasers to sense the differential phase accumulated between them as they propagate through the cavity by measuring their interference beatnote before their injection to the cavity and in transmission of it. With these measurements we were able to mitigate the effects of noise in our control loops in post-processing by up to a factor of 30 in the frequency range from 1\,mHz to 100\,Hz while suppressing injected calibration signals by over two orders of magnitude.
\end{abstract*}

\section{Introduction \label{Sec:Intro}}

Precision laser frequency stabilization systems that use optical cavities as a reference are critical to a diverse array of experiments such as optical clocks, gravitational wave observatories, and interferometric probes of new physics 
\cite{CLOCKLudlow2015, Aasi_2015, VIRGOAcernese2015, KAGRAAso2013, 3DTABLEVermeulen2021, ADBCPandey2024, ATOMBuchmueller2023}. 
Cavity frequency stabilization techniques can be combined with heterodyne sensing to investigate a range of other phenomena, from the optical properties of materials to the fundamental structure of the universe \cite{PhysRevA.108.013521,spector2026demonstration,PhysRevLett.132.191002}. Here, multiple lasers can be stabilized to different cavity resonances to observe how the spacing between cavity modes change due to some noise within the cavity or an external perturbation of it. While such methods can be powerful, they are vulnerable to sensing noise that will detune the laser frequencies from the cavity resonances. 
In this work, we demonstrate a heterodyne-enhanced sensing scheme that can measure this detuning and remove its effects in post-processing. 


One of the most well known laser frequency stabilization techniques is the Pound-Drever-Hall (PDH) method %
\cite{pound1946electronic,drever1983laser,black2001introduction}. 
PDH locking uses an electro-optic modulator (EOM) to induce phase modulation sidebands on a laser field in order to probe its resonance condition with respect to a cavity. As the laser field scans over a resonance, the carrier field in reflection of the cavity will experience a phase shift with respect to the sidebands, which are deliberately off-resonance. In reflection of the cavity, the beat of the sidebands with the carrier field leads to amplitude modulation on a photodetector, at the modulation frequency of the EOM. An error signal for locking the laser can be formed by demodulating the photodetector signal measured in reflection of the cavity at this frequency. 
Sensing noise, due to a number of different sources, can limit the precision of PDH locking. This includes residual amplitude modulation of the laser field as it passes through the EOM, electronic noise at the error point, or stray-light and parasitic etalons in the optical path coupling to the photodetector used for locking \cite{Zhang:14,Shi2018,RAManalysis}. 

The heterodyne-enhanced sensing scheme we propose is able to measure fluctuations in the differential locking error between two lasers by comparing the phase of their interference beatnote both before they are injected to the cavity and in transmission of it. The same cavity-induced phase shift that the PDH technique takes advantage of to sense the laser resonance condition in reflection of the cavity, can also be probed by the interference beatnote in transmission.
Any differential errors in the locking point between the two lasers, whether it is because of sensing or control noise or the finite gain of the loop, will lead to a differential phase between the beatnote measured at these points. The differential phase noise can be scaled to a differential frequency error in the PDH control loops using only the cavity storage time. Subtracting this frequency error from the measured frequency of the transmitted beatnote in post-processing allows the reference noise, or the changes in the frequency spacing between resonances of the cavity, to be more accurately resolved.



\section{Theory \label{Sec:Theory}}

To illustrate how the locking errors can be measured using the phase of the interference beatnotes, we will consider two laser fields $a$ and $b$.  These fields will be expressed as a superposition of Hermite-Gauss modes, $|\psi_{jk}\rangle$, of an optical cavity with angular frequencies $\omega_a$ and $\omega_b$ respectively:
\begin{align}
|\Psi_a\rangle &= e^{i\omega_a t}\left(a_{00}|\psi_{00}\rangle +a_{01}|\psi_{01}\rangle + a_{10}|\psi_{10}\rangle+... \right)
\\
|\Psi_b\rangle &=e^{i\omega_b t}\left( b_{00}|\psi_{00}\rangle +b_{01}|\psi_{01}\rangle + b_{10}|\psi_{10}\rangle+... \right).
\end{align}
All coefficients $a_{jk}$ and $b_{jk}$ can be complex. 
For simplicity we will express the fields as a superposition between the component of the field in the fundamental mode $|\psi_{00}\rangle $ and the combined wave function of all of the cavities higher order modes $|\Psi_{_{\rm HOM}}\rangle$: 
\begin{align}
|\Psi_a\rangle &= e^{i\omega_a t}\left(a_{00}|\psi_{00}\rangle +|\Psi_{_{a,\rm HOM}}\rangle\right)
 \\
|\Psi_b\rangle &= e^{i\omega_b t}\left(b_{00}|\psi_{00}\rangle +|\Psi_{_{b,\rm HOM}}\rangle\right).
\end{align}
The laser fields are assumed to be stabilized to a resonance of the cavity such that, in the simplest case,
\begin{align}
\omega_{ab} &= \omega_a-\omega_b
\\
& = \Delta N(\omega_0+\delta\omega_0)+\delta\omega_{ab}
\end{align}
where $\omega_0=2\pi f_0$, with $f_0=c/2L$ being the cavity free spectral range (FSR) for a cavity of length $L$. Here, $\Delta N =N_a-N_b$ refers to the difference between the longitudinal mode numbers of the cavity resonances, $N_a$ and $N_b$. The reference noise, $\delta\omega_{0}$, represents the angular frequency changes of the cavity FSR. 
The term $\delta\omega_{ ab} = \delta\omega_a-\delta\omega_b$ represents the sensing noise or the angular frequency deviation in the beatnote that is introduced when the lasers are not perfectly held on resonance with $\delta\omega_a$ and $\delta\omega_b$ representing the individual deviations of the respective lasers. 
The power present before the in-coupling mirror of the cavity can be expressed by
\begin{equation}
    (\langle\Psi_a| +\langle\Psi_b|) (|\Psi_a\rangle +|\Psi_b\rangle) = 
    |\langle \Psi_a\rangle|^2+|\langle \Psi_b\rangle|^2 +2 \Re[\langle\Psi_b|\Psi_a\rangle ] .    
\end{equation}
The first two terms on the right hand side of the equation give the static power of each of the laser fields, while the third term expresses the interference beatnote between them. The cross term, $\langle\Psi_b|\Psi_a\rangle$, can be expanded to
\begin{equation}
    \langle\Psi_b|\Psi_a\rangle =e^{i(\Delta N (\omega_0+\delta\omega_0) +\delta\omega_{ab})t} \left(\sum_{j=0}^\infty \sum_{k=0}^\infty a_{jk}b^*_{jk}\right),
    \label{Eq:cross_I}
\end{equation}
If both lasers are held directly on their corresponding resonance, the angular frequency of the beatnote should be exactly $\Delta N \omega_0$. 

The double sum in the previous equation can be simplified to the product of the fundamental mode coefficients and the product of the higher-order mode coefficients between the two lasers:
\begin{equation}
\sum_{j=0}^\infty \sum_{k=0}^\infty a_{jk}b^*_{jk} = a_{00} b_{00}^* +\langle\Psi_{_{b,\rm HOM}}|\Psi_{_{a,\rm HOM}}\rangle.
\end{equation}
The interference beatnote at the input of the cavity can therefore be expressed as 
\begin{multline}
2 \Re[\langle\Psi_b|\Psi_a\rangle ] = 2|a_{00}b^*_{00}|\cos((\Delta N (\omega_0+\delta\omega_0) +\delta\omega_{ab})t +\phi_{_{ab,00}})   \\
+A_{_{ab,{\rm HOM}}}\cos((\Delta N (\omega_0+\delta\omega_0) +\delta\omega_{ab})t+\phi_{_{ab,{\rm HOM}}}). 
\label{Eq:I_Re}    
\end{multline}
Here, $\phi_{_{ab,00}}$ represents the phase of the contribution of the fundamental mode to the interference beatnote, while $\phi_{_{ab,{\rm HOM}}}$ represents the same quantity due to the contribution of the higher-order modes. $A_{_{ab,{\rm HOM}}}$ is the amplitude of the higher-order mode component of the interference beatnote.

The interference beatnote in transmission of the cavity will be dependent on the real part of the cross term between fields $a$ and $b$ after applying the cavity transmission operator $\hat T$. It can be expressed as 
\begin{equation}
    \Re[\langle\Psi_b|\hat T^\dagger \hat T|\Psi_a\rangle] = \Re[e^{i(\Delta N (\omega_0+\delta\omega_0) +\delta\omega_{ab})t} a_{00} b_{00}^*\mathcal T(\delta\omega_a)\mathcal T^*(\delta\omega_b)],
    \label{Eq:cross_T}
\end{equation}
with $\mathcal T$ representing the transfer function of the cavity given by
\begin{align}
\mathcal T(\delta \omega) &=\frac{\sqrt{T_{\!_{1}}T_{\!_{2}}}}{1-\sqrt{1-l}\,e^{i \frac{\delta\omega}{f_0}}}\\
&\simeq \frac{2\sqrt{T_{\!_{1}}T_{\!_{2}}}}{l}\frac{1}{1-i\,\tau\delta\omega}.
\label{Eq:T_simp_rc}
\end{align}
In this equation, $l$ represents the total roundtrip optical losses in the cavity and $\tau$ represents the $1/e^2$ storage time of the cavity given by $2/lf_0$. In Equation~\ref{Eq:cross_T}, terms related to higher-order modes are ignored as high-finesse cavities act as a spatial filter for the transmitted field.

The expression for the beatnote in transmission then becomes
\begin{equation}
2 \Re[\langle\Psi_b|\hat T^\dagger \hat T|\Psi_a\rangle ]  =  A_{_{ab,\mathcal T}}\cos((\Delta N (\omega_0+\delta\omega_0)  +\delta\omega_{ab})t +\phi_{_{ab,00}}+\delta\phi_{_{ab,\mathcal T}})  .
\label{Eq:T_Re}
\end{equation}
The term  $A_{_{ab,\mathcal T}} = 2|\mathcal T(\delta\omega_a)||\mathcal T(\delta\omega_b)||a_{00}b^*_{00}|$ represents the amplitude of the beatnote while $\phi_{_{ab,\mathcal T}}={\rm arg}[\mathcal T(\delta\omega_a)\mathcal T^*(\delta\omega_b)]$ is the differential phase of the cavity transfer function for the two laser frequencies. This is due to their relative error with respect to the relevant cavity resonance and can be approximated by 
\begin{align}
    \delta\phi_{_{ab,\mathcal T}} &= \arctan{\tau\delta\omega_{ab}}\\
&\simeq \tau\delta\omega_{ab} \hspace{2cm} {\rm for ~}\tau\delta\omega_{a}{\rm ~and ~}\tau\delta\omega_{b}\ll1.
\end{align}
The difference in the phase of the transmitted beatnote from Equation~\ref{Eq:T_Re} and the phase of the fundamental mode contribution to the beatnote at the input of the cavity from Equation~\ref{Eq:I_Re} is equal to $\phi_{_{ab,\mathcal T}}$. Therefore, if the fundamental mode contribution to the input beatnote can be distinguished from the higher order mode contributions, $\phi_{_{ab,\mathcal T}}$, and hence the different sensing errors in the PDH loops $\delta\omega_{_{ab}}$, can be measured independent of the reference noise $\delta\omega_0$, that is due to the length changes of the cavity.

If the lasers occupy the same spatial eigenmode the higher order mode coefficients in Equation~\ref{Eq:I_Re} will have identical phases and because of this their product in Equation~\ref{Eq:cross_I}, $a_{jk}b^*_{jk}$, will be real valued.  Therefore, in Equation~\ref{Eq:I_Re}, $\phi_{_{ab,00}}$ will be equal to $\phi_{_{ab,{\rm HOM}}}$. If this is the case, then the differential phase between the input and transmitted beatnotes can be expressed as
\begin{equation}
   \phi_t-\phi_i =  \phi_{_{ab,\mathcal T}} +\Delta\phi_{_{ab,\rm PD}},
\end{equation}
where $\phi_t$ is the phase of the beatnote measured in transmission of the cavity, $\phi_i$ is the phase measured before incidence on the cavity.  The static phase difference between the beatnotes due to phase delays from features in the system such as differences in cable lengths for the signal or the optical path length between the cavity mirrors and the photodetectors is represented by $\Delta\phi_{_{ab,\rm PD}}$. Changes in such parameters have a negligible effect on the measured phase as their coupling scales inversely with the wavelength of the beatnote which typically is on the order of the cavity length.

As $\Delta\phi_{_{ab,\rm PD}}$ is static, the difference in the dynamic fluctuations between the transmitted and incident beatnote phases, given by $\delta\phi_t=\phi_t-\langle\phi_t\rangle$ and  $\delta\phi_i=\phi_i-\langle\phi_i\rangle$, can be used to calculate the dynamic differential frequency error of the fields with respect to their cavity resonances with 
\begin{equation}
   \delta\omega_{ab} = \frac{\delta\phi_t-\delta\phi_i} {\tau}.
   \label{Eq:del_om}
\end{equation}
To more accurately resolve the frequency changes between the cavity resonances, $\delta\omega_{ab}$ can be subtracted from the time series of the beatnote frequency measured in transmission of the cavity $\delta f_t$:
\begin{equation}
   \delta f_{\rm cav} = \delta f_t - \frac{\delta\phi_t-\delta\phi_i} {2\pi\tau}.
   \label{Eq:true_FSR}
\end{equation}
Here, $\delta f_{\rm cav}$ represents the measured changes in the frequency spacing between the resonances. 
In the ideal case where the lasers are stabilized to cavity resonances which only differ in their longitudinal mode order, $\delta f_{\rm cav} = \Delta N\delta f_0$, with $\delta f_0=\delta\omega_0/2\pi$ giving the changes in the FSR of the cavity.




\section{Methods \label{Sec:Methods}}

\begin{figure}
    \centering
\includegraphics[width=0.75\linewidth]{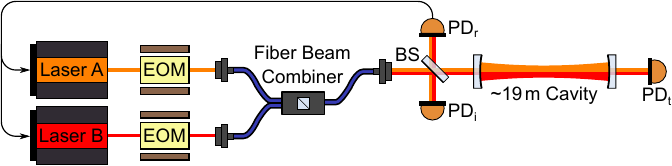}
    \caption{The optical system for the experiment. Two lasers provide the fields $a$ (shown in orange) and  $b$ (shown in red). 
}
    \label{fig:Proto}
\end{figure}

The heterodyne-enhanced sensing technique was tested using a 19\,m optical cavity. A detailed description of the cavity is available in Refs.~\cite{spector2026demonstration} and \cite{Spector:24}. Figure~\ref{fig:Proto} shows the layout of the optical system. The fields $a$ and $b$ were generated with two Nd:YAG non-planar ring oscillators operating at a wavelength of 1064\,nm (labeled Laser A and Laser B in the figure). Both fields passed through EOMs that introduced phase modulation sidebands used to sense their resonance condition with respect to the cavity at the photodetector $\rm PD_r$ via the PDH technique. The PDH error signal was sent to a PID controller and fed back to both piezo and temperature actuators mounted to the laser crystals. The fields were stabilized to resonances separated by one FSR of the cavity and produced a beatnote frequency of $f_0 = 7.89165\pm0.00003$\,MHz. The beatnote frequency could be measured with sub-hertz precision, but drifted by roughly 30\,Hz due to the length changes of the cavity over the course of the measurement. During the measurements the storage time of the cavity was $\tau=1.308\pm0.010\rm\,ms$ and the spatial overlap between the lasers and the fundamental mode of the cavity was  $97\pm1\%$.

The photodetector $\rm PD_i$ sensed the interference beatnote before it was incident on the cavity while the photodetector $\rm PD_t$ sensed the beatnote in transmission. The electronic signals from both photodetectors were sent to a digital phasemeter that was used to readout their phase and frequency with respect to a rubidium standard clock.




\section{Results \label{Sec:Results}}

\begin{figure}
    \centering
    \includegraphics[width=0.75\linewidth]{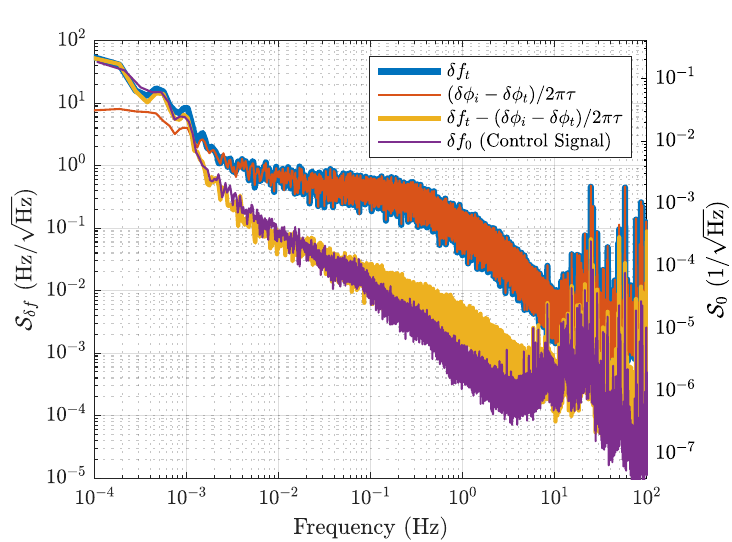}
    \caption{Measured frequency noise in the interference beatnote of the two lasers stabilized to the cavity. The $y$-axis on the left is in units of $\rm Hz/\sqrt{Hz}$ while the $y$-axis on the right is scaled as a fractional error with respect to the cavity FWHM. 
    }
    \label{fig:results}
\end{figure}

The results can be seen in Figure~\ref{fig:results} with the $x$-axis representing the Fourier frequency, the $y$-axis on the left representing frequency noise amplitude spectral density in $\rm Hz/\sqrt{Hz}$, and the $y$-axis on the right representing frequency noise amplitude spectral density normalized by the FWHM of the cavity (246\,Hz) in units of $\rm 1/\sqrt{Hz}$. The blue trace in this plot shows the frequency noise of the interference beatnote, or $\delta f_t$, measured at $\rm PD_t$. The orange trace shows $\delta\omega_{ab}/2\pi$ from Equation~\ref{Eq:del_om}, or the differential frequency detuning between the lasers and their respective cavity resonances. The yellow trace is $\delta f_{\rm cav}$ from Equation~\ref{Eq:true_FSR}, which ideally is equivalent to the changes in the cavity FSR $\delta f_0$, as $\Delta N=1$. This was found by subtracting $\delta\omega_{ab}/2\pi$ from $\delta f_t$ to mitigate the effects of the detuning and more precisely resolve the changes in the frequency difference between the cavity resonances. The purple trace shows a scaled spectrum of the PDH control signal acting on the frequency of Laser A while it was stabilized to the cavity. At the shown frequencies the control signal can be used to estimate the changes in the FSR, $\delta f_0$, since the length noise of the 19\,m cavity in terms of frequency is typically higher than the free-running laser frequency noise \cite{spector2026demonstration,Spector:24}. The spectrum of the frequency noise in the cavity FSR signal is found using the the equation $\frac{\delta f_0}{f_0}=\frac{\delta\nu}{\nu}$, where $\nu$ represents the frequency of the laser locked to the cavity and $\delta\nu$ its frequency noise due to the cavity length noise. 

As is seen in the plot, above 1\,mHz the measured frequency detuning (orange) shows exceptional agreement with the $\delta f_t$ (blue). Below 1\,mHz the cavity length noise (purple) then becomes the dominant source of the beat signal's frequency noise. This demonstrates that for most of the spectra, the leading cause of the frequency noise in the beatnote is the residual noise in the laser frequency stabilization systems. 


The corrected frequency spectra shown in yellow illustrates the capabilities of this technique as the estimate of the residual differential frequency noise was suppressed in post processing by nearly an order of magnitude between 10\,mHz and 10\,Hz and up to a factor of 30 at some frequencies over this range. Furthermore, below 10\,mHz and from 8\,Hz to 50\,Hz the measurement of $\delta f_{\rm cav}$ (yellow) was able to resolve the frequency changes between the cavity resonances $\delta f_0$ (purple) that are due to the length noise of the cavity. 

It is clear that from roughly 100\,mHz to 5\,Hz, there is excess noise which limits the sensitivity of $\delta f_{\rm cav}$, as its spectra is above the cavity length noise, $\delta f_0$ . While the exact origin of this noise is unknown, it is believed to be due to stray-light creating small vector phase noise at the photodetectors. 
Nevertheless, comparing the sensitivity of the corrected to the uncorrected beatnote frequency noise spectra, the corrected data is clearly able to resolve the cavity length noise for integration times less than 100\,s, while the uncorrected frequency data requires at least 1000\,s to do so.


\begin{figure}
    \centering
    \includegraphics[width=0.75\linewidth]{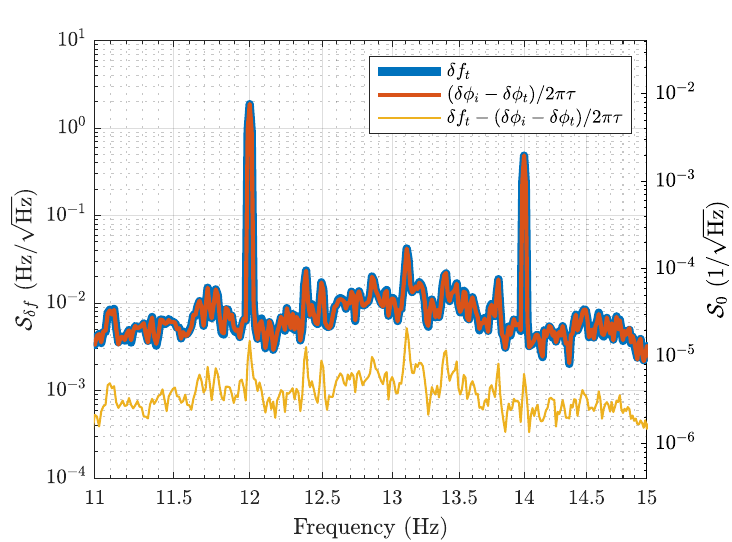}
    \caption{Results from the measurement of the noise in the beatnote frequency with 12\,Hz and 14\,Hz signals injected to the error points of the laser A and laser B frequency stabilization systems respectively. 
    }
    \label{fig:line}
\end{figure}

Another demonstration of the technique is shown in Figure~\ref{fig:line} with the colors corresponding to the same spectra as in Figure~\ref{fig:results}. Here, sinusoids were injected to the error points of the Laser A and Laser B PDH locks at 12\,Hz and 14\,Hz respectively. The amplitude of these signals were well above the out-of-loop noise of the frequency stabilization systems. 
In these spectra, the lines are clearly seen in $\delta f_t$ (blue) and the frequency detuning of the lasers from the cavity resonances (orange), however, when the corrected beatnote frequency $\delta f_{\rm cav}$ (yellow) is calculated, the lines nearly disappear. 
This shows that the technique is capable of reducing the estimated residual frequency noise by more than two orders of magnitude at these frequencies. 


\section{Conclusion}

We have shown that heterodyne-enhanced transmission sensing is capable of distinguishing between the differential noise of two lasers with respect to their cavity resonances (corresponding to the quality of the frequency stabilization), and the frequency changes between the resonances themselves. The ability to perform out-of-loop measurements of the differential noise in the laser frequency control loops (the orange traces in Figs.~\ref{fig:results} and~\ref{fig:line}) makes the technique a valuable  diagnostic tool that can be used to characterize laser frequency stabilization systems. Furthermore, encoding this noise on the phase of a MHz beatnote, due to the use of heterodyne sensing, makes it a robust method for measuring low-frequency error point noise.

Going beyond the diagnostic capabilities of the technique, the effects of the additional noise in the system can be removed in post processing to improve the sensing precision of the cavity mode spacing. While in this work we have only explored using heterodyne-enhanced sensing to measure the difference between longitudinal modes of identical polarization states, this technique can be expanded to also investigate other mode spacings, such as those between higher-order modes or polarization eigenmodes. This is especially relevant for experiments that rely on the frequency stabilization of lasers to optical cavities and may even be used to increase the sensitivity of interferometric probes of fundamental physics \cite{Kedar:23,PhysRevA.62.013815,spector2026demonstration}. In fact, it should be possible to apply the principles of heterodyne-enhanced sensing more broadly to investigate resonant media, even absent the use of an optical cavity, by measuring the interference beatnote between two lasers before and after passing through a medium that exhibits resonant modes. 

\begin{backmatter}
\bmsection{Acknowledgment} We thank the ALPS Collaboration for their support, especially Todd Kozlowski for his work on the laboratory infrastructure, as well as Guido Müller, Benno Willke, Hartmut Grote, Aldo Ejlli, David B. Tanner, and Axel Lindner for the many interesting discussions and feedback on the manuscript. We also thank Heinrich Pryschelski and Frank Ludwig of the MSK group at DESY for their consultation on timing noise and lending us equipment for the experiment. 
\bmsection{Funding} The work is supported by the Deutsche Forschungsgemeinschaft under Germany’s Excellence Strategy – EXC 2121 ``Quantum Universe" – 390833306 and - grant number WI 1643/2-1, the Partnership for Innovation, Education and Research (PIER) of DESY and Universit\"at Hamburg under PIER Seed Project - PIF-2022-18, the National Science Foundation  - grant numbers PHY-2110705 365 and PHY-1802006, the Heising Simons Foundation - grant numbers 2015-154 and 2020-1841. Laura Roberts acknowledges support by grant number 63132 from the John Templeton Foundation, as recipient of an Enrico Fermi Fellowship awarded through the Center for SpaceTime and the Quantum.
\end{backmatter}

\bibliography{het_phi_corr}

\begin{thebibliography}{10}
\newcommand{\enquote}[1]{``#1''}

\bibitem{CLOCKLudlow2015}
A.~D. Ludlow, M.~M. Boyd, J.~Ye, \emph{et~al.}, \enquote{Optical atomic clocks,} {\protect\JournalTitle{Rev. Mod. Phys.}} \textbf{87}, 637--701 (2015).

\bibitem{Aasi_2015}
J.~Aasi, B.~P. Abbott, R.~Abbott, \emph{et~al.}, \enquote{Advanced {LIGO},} {\protect\JournalTitle{Classical and Quantum Gravity}} \textbf{32}, 074001 (2015).

\bibitem{VIRGOAcernese2015}
F.~Acernese, M.~Agathos, K.~Agatsuma, \emph{et~al.}, \enquote{Advanced {Virgo}: a second-generation interferometric gravitational wave detector,} {\protect\JournalTitle{Classical and Quantum Gravity}} \textbf{32}, 024001 (2014).

\bibitem{KAGRAAso2013}
Y.~Aso, Y.~Michimura, K.~Somiya, \emph{et~al.}, \enquote{Interferometer design of the {KAGRA} gravitational wave detector,} {\protect\JournalTitle{Phys. Rev. D}} \textbf{88}, 043007 (2013).

\bibitem{3DTABLEVermeulen2021}
S.~M. Vermeulen, L.~Aiello, A.~Ejlli, \emph{et~al.}, \enquote{An experiment for observing quantum gravity phenomena using twin table-top {3D} interferometers,} {\protect\JournalTitle{Classical and Quantum Gravity}} \textbf{38}, 085008 (2021).

\bibitem{ADBCPandey2024}
S.~Pandey, E.~D. Hall, and M.~Evans, \enquote{First results from the {Axion Dark-Matter Birefringent Cavity (ADBC)} experiment,} {\protect\JournalTitle{Phys. Rev. Lett.}} \textbf{133}, 111003 (2024).

\bibitem{ATOMBuchmueller2023}
O.~Buchmueller, J.~Ellis, and U.~Schneider, \enquote{Large-scale atom interferometry for fundamental physics,} {\protect\JournalTitle{Contemporary Physics}} \textbf{64}, 93--110 (2023).

\bibitem{PhysRevA.108.013521}
N.~Cooper, S.~Madkhaly, D.~Johnson, \emph{et~al.}, \enquote{Dual-frequency {Doppler}-free spectroscopy for simultaneous laser stabilization in compact atomic physics experiments,} {\protect\JournalTitle{Phys. Rev. A}} \textbf{108}, 013521 (2023).

\bibitem{spector2026demonstration}
A.~D. Spector, T.~Kozlowski, and L.~Roberts, \enquote{Demonstration of an interferometric technique for measuring vacuum magnetic birefringence with an optical cavity,} {\protect\JournalTitle{Physical Review A}} \textbf{113}, 063518 (2026).

\bibitem{PhysRevLett.132.191002}
J.~Heinze, A.~Gill, A.~Dmitriev, \emph{et~al.}, \enquote{First results of the {Laser-Interferometric Detector for Axions (LIDA)},} {\protect\JournalTitle{Phys. Rev. Lett.}} \textbf{132}, 191002 (2024).

\bibitem{pound1946electronic}
R.~V. Pound, \enquote{Electronic frequency stabilization of microwave oscillators,} {\protect\JournalTitle{Rev. Sci. Instrum.}} \textbf{17}, 490--505 (1946).

\bibitem{drever1983laser}
R.~W.~P. Drever, J.~L. Hall, F.~Kowalski, \emph{et~al.}, \enquote{Laser phase and frequency stabilization using an optical resonator,} {\protect\JournalTitle{Appl. Phys. B}} \textbf{31}, 97--105 (1983).

\bibitem{black2001introduction}
E.~D. Black, \enquote{An introduction to {Pound--Drever--Hall} laser frequency stabilization,} {\protect\JournalTitle{Am. J. Phys.}} \textbf{69}, 79--87 (2001).

\bibitem{Zhang:14}
W.~Zhang, M.~J. Martin, C.~Benko, \emph{et~al.}, \enquote{Reduction of residual amplitude modulation to $1\times10^{-6}$ for frequency modulation and laser stabilization,} {\protect\JournalTitle{Opt. Lett.}} \textbf{39}, 1980--1983 (2014).

\bibitem{Shi2018}
X.~Shi, J.~Zhang, X.~Zeng, \emph{et~al.}, \enquote{Suppression of residual amplitude modulation effects in {Pound--Drever--Hall} locking,} {\protect\JournalTitle{Applied Physics B}} \textbf{124}, 153 (2018).

\bibitem{RAManalysis}
H.~Shen, L.~Li, J.~Bi, \emph{et~al.}, \enquote{Systematic and quantitative analysis of residual amplitude modulation in {Pound--Drever--Hall} frequency stabilization,} {\protect\JournalTitle{Phys. Rev. A}} \textbf{92}, 063809 (2015).

\bibitem{Spector:24}
A.~D. Spector and T.~Kozlowski, \enquote{Optical cavity characterization with a mode-matched heterodyne sensing scheme,} {\protect\JournalTitle{Opt. Express}} \textbf{32}, 27112--27124 (2024).

\bibitem{Kedar:23}
D.~Kedar, J.~Yu, E.~Oelker, \emph{et~al.}, \enquote{Frequency stability of cryogenic silicon cavities with semiconductor crystalline coatings,} {\protect\JournalTitle{Optica}} \textbf{10}, 464--470 (2023).

\bibitem{PhysRevA.62.013815}
J.~L. Hall, J.~Ye, and L.-S. Ma, \enquote{Measurement of mirror birefringence at the sub-ppm level: {Proposed} application to a test of {QED},} {\protect\JournalTitle{Phys. Rev. A}} \textbf{62}, 013815 (2000).

\end{thebibliography}
\end{document}